\documentclass{epl}
\newcommand{\half}{{\textstyle \frac{1}{2}}}
\def\hbarit {{\mathchar'26\mkern-11muh}} 
\title{Fano diagonalization of a polariton model for
an inhomogeneous absorptive dielectric}
\shorttitle{Fano diagonalization}
\author{L.G.\ Suttorp\inst{1} and A.J.\ van Wonderen\inst{1}} 
\institute{ \inst{1} Institute for Theoretical Physics,
University of Amsterdam, Valckenierstraat 65, NL-1018 XE Amsterdam, The
Netherlands } 
\pacs{3.70.+k}{Theory of quantized fields}
\begin{document}
\maketitle

\begin{abstract}
The Hamiltonian of a polariton model for an inhomogeneous linear absorptive
dielectric is diagonalized by means of Fano's diagonalization method. The
creation and annihilation operators for the independent normal modes are
explicitly found as linear combinations of the canonical operators. The
coefficients in these combinations depend on the tensorial Green function
that governs the propagation of electromagnetic waves through the
dielectric. The time-dependent electromagnetic fields in the Heisenberg
picture are given in terms of the diagonalizing operators. These results
justify the phenomenological quantization of the electromagnetic field in
an absorptive dielectric.
\end{abstract}

\section{Introduction}
To describe quantum optical phenomena in dielectrics it is essential to
have available a quantization procedure for the electromagnetic field in
ponderable matter. Preferably, such a quantization should be based on the
standard canonical quantization method of quantum field theory. To
apply that method to the fields in linear dielectrics, one should start
from a Hamiltonian description, in which both the field and the dielectric
are given in terms of canonical variables that are coupled bilinearly. The
normal modes in such a system are the well-known polaritons~\cite{H58}. If
the dielectric is absorptive and dispersive, damping of the polariton modes
may be taken into account by coupling the dielectric degrees of freedom to
a suitable bath of harmonic oscillators with a continuous range of
frequencies. Working along these lines, Huttner and Barnett~\cite{HB92}
were the first to formulate a damped-polariton model for an absorptive
dielectric in an electromagnetic field and to study its properties. By
using a diagonalization method due to Fano~\cite{F61} they were able to
find the full time dependence of the electromagnetic field operators for
their model.

The method of diagonalization employed in~\cite{HB92} is based on a
separation of longitudinal and transverse degrees of freedom, and on a
Fourier decomposition of the canonical variables. Both of these means are
only expedient for homogeneous systems with translation invariance. In
fact, when the damped-polariton model is taken to be inhomogeneous, with an
arbitrary spatial dependence of the material properties, the longitudinal
and transverse degrees of freedom get coupled, while the Fourier components
of the canonical variables for different wave vectors start interacting as
well. Hence, the diagonalization procedure in~\cite{HB92} runs into
difficulties for the inhomogeneous case. Since inhomogeneities are
unavoidable in any quantum optical experiment involving dielectrics, this
is a serious drawback. It is the purpose of this Letter to show that
diagonalization of the inhomogeneous version of the damped-polariton model
is possible, and that explicit expressions for the diagonalizing operators
in terms of canonical variables can be found.

\section{Model}
In the damped-polariton model the polarization density is coupled to a bath
of harmonic oscillators with a continuous range of eigenfrequencies. The
electromagnetic field interacts with the dielectric according to the
standard minimal-coupling scheme. The Hamiltonian of the model
is~\cite{HB92}
\begin{eqnarray}
H&=&\int \upd {\bm r}\left[\frac{1}{2\epsilon_0} \Pi^2+
\frac{1}{2 \mu_0} ({\bm \nabla}\times {\bm A})^2 
+\frac{1}{2\rho}\, P^2+\half \rho\tilde{\omega}_0^2\, X^2 
+\frac{1}{2\rho}\int_0^{\infty}\upd \omega\, Q_{\omega}^2 
\right. \nonumber\\
&&\left.+\half\rho\int_0^{\infty}\upd\omega\, \omega^2\, Y_{\omega}^2
 +\frac{\alpha}{\rho}{\bm A}\cdot{\bm P}+\frac{\alpha^2}{2\rho}\, A^2
+\frac{1}{\rho}\int_0^{\infty}\upd\omega\, v_{\omega}\, {\bm X}\cdot{\bm
Q}_{\omega}\right]\nonumber\\
&&+\int\upd{\bm r}\upd{\bm r}'\,
\frac{{\bm \nabla}\cdot(\alpha{\bm X})\, {\bm \nabla'}\cdot(\alpha'{\bm X}')}
{8\pi\epsilon_0|{\bm r}-{\bm r}'|}\, .\label{2.1}
\end{eqnarray}
The transverse part of the electromagnetic field is determined by the
vector potential ${\bm A}({\bm r})$, for which the Coulomb gauge is
adopted. Its conjugate canonical momentum is ${\bm \Pi}({\bm r})$. The
linear dielectric, with a space-dependent density $\rho({\bm r})$, is
described by the harmonic displacement variable ${\bm X}({\bm r})$ and its
canonical momentum ${\bm P}({\bm r})$. The associated (renormalized)
eigenfrequency $\tilde{\omega}_0({\bm r})$ is generally space-dependent as
well. The electromagnetic field is coupled to the dielectric variable ${\bm
X}$ in the usual way. In terms of the polarization density $-\alpha{\bm
X}$, with a space-dependent coupling parameter $\alpha({\bm r})>0$, the
minimal coupling scheme leads to an electrostatic contribution and to a
bilinear interaction term with ${\bm A}\cdot{\bm P}$. Finally, damping is
introduced in the system by a continuum bath of harmonic oscillators with
canonical variables ${\bm Y}_{\omega}({\bm r})$, ${\bm Q}_{\omega}({\bm
r})$ and with eigenfrequencies $\omega$. These bath oscillators are coupled
to ${\bm X}({\bm r})$ with a strength $v_{\omega}({\bm r})>0$. We used the
notation ${\bm X}'={\bm X}({\bm r}')$, and likewise $\alpha'$ and ${\bm
\nabla'}$.

The canonical variables obey the standard commutation relations
\begin{eqnarray}
\left[{\bm \Pi}({\bm r}),{\bm A}({\bm r}')\right]& = & 
-i\,\hbarit\, {\bm \delta}_\mathrm{T}({\bm r}-{\bm r}')\, , \qquad 
\left[{\bm P}({\bm r}),{\bm X}({\bm r}')\right] =  
-i\,\hbarit\, \tens{I} \, \delta({\bm r}-{\bm r}')\, ,\nonumber\\ 
\left[{\bm Q}_{\omega}({\bm r}),{\bm Y}_{\omega'}({\bm r}')\right]& = &
-i\,\hbarit\, \delta(\omega-\omega')\, \tens{I} \,\delta({\bm r}-{\bm
r}')\, ,
\label{2.2}
\end{eqnarray}
while all other commutators of the canonical variables vanish. Here
$\tens{I}$ is the three-dimensional unit tensor, while ${\bm
\delta}_\mathrm{T}({\bm r})=\tens{I}\, \delta({\bm r})
+{\bm\nabla}{\bm\nabla}(4\pi r)^{-1}$ is the transverse delta function.

The electric field operator ${\bm E}$ is the sum of a transverse
part depending on ${\bm \Pi}$ and a longitudinal part that is proportional
to the polarization density:
\begin{equation}
{\bm E}({\bm r})=-\frac{1}{\varepsilon_0}\, {\bm \Pi}({\bm
r})+\frac{1}{\varepsilon_0}\, [\alpha{\bm X}({\bm r})]_\mathrm{L}\, .
\label{2.3}
\end{equation}
The longitudinal part of a vector (or a tensor) is obtained by a
convolution with the longitudinal delta function ${\bm \delta}_\mathrm{L}
({\bm r})=-{\bm\nabla}{\bm\nabla}(4\pi r)^{-1}$. The displacement field 
\begin{equation}
{\bm D}({\bm r})=-{\bm \Pi}({\bm r})-[\alpha{\bm X}({\bm r})]_\mathrm{T}\, 
\label{2.4}
\end{equation}
is purely transverse. 

In the following we will show how the Hamiltonian (\ref{2.1}), with
canonical operators satisfying the commutation relations (\ref{2.2}), can
be brought in diagonal form.

\section{Fano diagonalization}
The Hamiltonian is quadratic in the canonical variables. Hence, it should
be possible to find a diagonal representation of the form
\begin{equation}
H=\int \upd {\bm r}\int_0^{\infty}\upd\omega\, \hbarit\omega \, {\bm
C}^{\dagger}({\bm r},\omega)\cdot {\bm C}({\bm r},\omega)\, , \label{3.1}
\end{equation}
where we omit a zero-point-energy term. The operators ${\bm C}({\bm
r},\omega)$ are annihilation operators, which (together with the associated
creation operators) satisfy the commutation relations:
\begin{equation}
\left[{\bm C}({\bm r},\omega),{\bm C}^{\dagger}({\bm r}',\omega')\right]=
\delta(\omega-\omega')\, \tens{I} \, \delta({\bm r}-{\bm r}')\, , \qquad
\left[{\bm C}({\bm r},\omega),{\bm C}({\bm r}',\omega')\right]=
0\, .
\label{3.2}
\end{equation}

Each canonical operator can be written as a linear combination of
the annihilation and creation operators. For instance, one has
\begin{eqnarray}
{\bm A}({\bm r})&=&\int\upd{\bm r}'\int_0^{\infty}\upd\omega \, 
\tens{f}_A({\bm r},{\bm r}',\omega)\cdot{\bm C}({\bm r}',\omega) 
+\tx{h.c.}\, , \nonumber\\
{\bm Q}_{\omega}({\bm r})&=&\int\upd{\bm r}'\int_0^{\infty}\upd\omega' \, 
\tens{f}_Q({\bm r},{\bm r}',\omega,\omega')
\cdot{\bm C}({\bm r}',\omega') +\tx{h.c.}\, , \label{3.3}
\end{eqnarray}
with tensorial coefficients $\tens{f}_A$ and $\tens{f}_Q$. The coefficients
$\tens{f}_{\Pi}$, $\tens{f}_X$, $\tens{f}_P$ and $\tens{f}_Y$ are defined
analogously. Both $\tens{f}_A$ and $\tens{f}_{\Pi}$ are transverse in ${\bm
  r}$. The electric field follows from (\ref{2.3}) as
\begin{equation}
{\bm E}({\bm r})=\int\upd{\bm r}'\int_0^{\infty}\upd\omega \, 
\tens{f}_E({\bm r},{\bm r}',\omega)\cdot{\bm C}({\bm r}',\omega) 
+\tx{h.c.}\, , \label{3.4}
\end{equation}
with the coefficient 
\begin{equation}
\tens{f}_{E}({\bm r},{\bm r}',\omega)=
-\frac{1}{\varepsilon_0}\, \tens{f}_{\Pi}({\bm r},{\bm r}',\omega)+
\frac{1}{\varepsilon_0}\, \left[\alpha\, \tens{f}_{X}({\bm r},{\bm
    r}',\omega)\right]_\mathrm{L}\, . \label{3.5}
\end{equation}

From eq.~(\ref{3.2}) it follows that the coefficients are equal
to commutators. For instance, one has:
\begin{equation}
\tens{f}_A({\bm r},{\bm r}',\omega)=\left[{\bm A}({\bm r}),{\bm
C}^{\dagger}({\bm r}',\omega)\right] \, , \qquad
\tens{f}_Q({\bm r},{\bm r}',\omega,\omega')=
\left[{\bm Q}_{\omega}({\bm r}),{\bm C}^{\dagger}({\bm r}',\omega')\right] 
\, . \label{3.6}
\end{equation}
Inversely, each ${\bm C}({\bm r},\omega)$ is a linear combination of the
canonical operators:
\begin{eqnarray}
{\bm C}({\bm r},\omega)&=&-\frac{i}{\hbarit}\, \int\upd{\bm r}'\, \biggl\{
{\bm A}({\bm r}')\cdot\tens{f}^{\ast}_{\Pi}({\bm r}',{\bm r},\omega)-
{\bm \Pi}({\bm r}')\cdot\tens{f}^{\ast}_{A}({\bm r}',{\bm r},\omega)
\nonumber\\
&&+{\bm X}({\bm r}')\cdot\tens{f}^{\ast}_{P}({\bm r}',{\bm r},\omega)
-{\bm P}({\bm r}')\cdot\tens{f}^{\ast}_{X}({\bm r}',{\bm r},\omega)
\nonumber\\
&&+\int_0^{\infty}\upd\omega'\,\left[
{\bm Y}_{\omega'}({\bm r}')\cdot\tens{f}^{\ast}_{Q}({\bm r}',{\bm
r},\omega',\omega)-
{\bm Q}_{\omega'}({\bm r}')\cdot\tens{f}^{\ast}_{Y}({\bm r}',{\bm
r},\omega',\omega)\right]\biggr\}\, . \label{3.7}
\end{eqnarray}
Fano's method to diagonalize the Hamiltonian amounts to finding the
tensorial coefficients in these expressions by solving a set of equations
that follows from the commutator of ${\bm C}$ with the Hamiltonian. In fact,
eqs.~(\ref{3.1}) and (\ref{3.2}) imply:
\begin{equation}
\left[{\bm C}({\bm r},\omega),H\right]=\hbarit\omega\, {\bm C}({\bm
r},\omega)\, .\label{3.8}
\end{equation}
Upon inserting eqs.~(\ref{2.1}) and (\ref{3.7}), employing (\ref{2.2}) and
comparing the coefficients of the canonical operators, we arrive at the
following set of linear relations:
\begin{eqnarray}
i\omega\, \tens{f}_{A}({\bm r},{\bm r}',\omega)&=&-\frac{1}{\varepsilon_0}\, 
\tens{f}_{\Pi}({\bm r},{\bm r}',\omega)\, , \label{3.9}\\
i\omega\, \tens{f}_{\Pi}({\bm r},{\bm r}',\omega)&=&-\frac{1}{\mu_0}\,
\Delta \, \tens{f}_{A}({\bm r},{\bm r}',\omega)
+\left[\frac{\alpha^2}{\rho}\, 
\tens{f}_{A}({\bm r},{\bm r}',\omega)\right]_\mathrm{T}
+\left[ \frac{\alpha}{\rho}\, \tens{f}_{P}({\bm r},{\bm
r}',\omega)\right]_\mathrm{T} \, , \label{3.10}\\
i\omega\, \tens{f}_{X}({\bm r},{\bm r}',\omega)&=&
-\frac{\alpha}{\rho}\, \tens{f}_{A}({\bm r},{\bm r}',\omega)
-\frac{1}{\rho}\, \tens{f}_{P}({\bm r},{\bm r}',\omega)\, , \label{3.11}\\
i\omega\, \tens{f}_{P}({\bm r},{\bm r}',\omega)&=&
\rho\tilde{\omega}_0^2\, \tens{f}_{X}({\bm r},{\bm r}',\omega)
+\frac{\alpha}{\varepsilon_0}\, \left[\alpha\, 
\tens{f}_{X}({\bm r},{\bm r}',\omega)\right]_\mathrm{L}\nonumber\\
&&+\frac{1}{\rho}\int_0^{\infty}\upd\omega'\, v_{\omega'}\,
\tens{f}_{Q}({\bm r},{\bm r}',\omega',\omega) \, , \label{3.12}\\
i\omega\, \tens{f}_{Y}({\bm r},{\bm r}',\omega',\omega)&=&
-\frac{1}{\rho}\, v_{\omega'}\, \tens{f}_{X}({\bm r},{\bm r}',\omega)
-\frac{1}{\rho}\, \tens{f}_{Q}({\bm r},{\bm r}',\omega',\omega)\, ,
\label{3.13}\\
i\omega\, \tens{f}_{Q}({\bm r},{\bm r}',\omega',\omega)&=&
\rho{\omega'}^2\, \tens{f}_{Y}({\bm r},{\bm r}',\omega',\omega)\,
, \label{3.14}
\end{eqnarray}
with $\alpha$, $\rho$, $\tilde{\omega}_0$ and $v_{\omega}$ depending on
${\bm r}$. To solve these equations we shall first consider the purely
algebraic equations~(\ref{3.9}), (\ref{3.11})--(\ref{3.14}). This will lead
to the introduction of the susceptibility.

\section{Introduction of the susceptibility}
By elimination of $\tens{f}_{Y}$ from eqs.~(\ref{3.13}) and (\ref{3.14}) we
get
\begin{equation}
\left(\omega^2-{\omega'}^2\right)\, 
\tens{f}_{Q}({\bm r},{\bm r}',\omega',\omega)={\omega'}^2v_{\omega'}\, 
\tens{f}_{X}({\bm r},{\bm r}',\omega) \, . \label{4.1}
\end{equation}
To obtain $\tens{f}_{Q}$ we have to impose a prescription for the pole at
$\omega=\omega'$:
\begin{equation}
\tens{f}_{Q}({\bm r},{\bm r}',\omega',\omega)=
\frac{{\omega'}^2\, v_{\omega'}}{(\omega+i0)^2-{\omega'}^2}\, 
\tens{f}_{X}({\bm r},{\bm r}',\omega)+
\tens{s}({\bm r},{\bm r}',\omega)\, \delta(\omega-\omega')\, , 
\label{4.2}
\end{equation}
with $\omega+i0$ a complex frequency in the upper half-plane and
infinitesimally close to the real axis. The last term contains an as yet
unknown tensor $\tens{s}({\bm r},{\bm r}',\omega)$. Substituting
eq.~(\ref{4.2}) in (\ref{3.12}), and eliminating $\tens{f}_{P}$ and
$\tens{f}_{A}$ with the help of eqs.~(\ref{3.9}) and (\ref{3.11}) we get an
equality which expresses $\tens{f}_{\Pi}$ in terms of $\tens{f}_{X}$ and
$\tens{s}$. If we use (\ref{3.5}) to eliminate $\tens{f}_{\Pi}$ in favor
of $\tens{f}_E$, this equality may be rewritten as a linear relation
between the coefficients of the polarization density $-\alpha{\bm X}$
and the electric field ${\bm E}$:
\begin{equation}
-\alpha\tens{f}_{X}({\bm r},{\bm r}',\omega)=\varepsilon_0\, 
\chi({\bm r},\omega)\, \tens{f}_{E}({\bm r},{\bm r}',\omega)
+\frac{\varepsilon_0}{\rho\alpha}\, v_{\omega}\, 
\chi({\bm r},\omega)\, \tens{s}({\bm r},{\bm r}',\omega)\, . 
\label{4.3}
\end{equation}
The proportionality constant is the position- and frequency-dependent
susceptibility:
\begin{equation}
\chi({\bm r},\omega)=-\frac{\alpha^2}{\varepsilon_0\rho}\, 
\left[\omega^2-\tilde{\omega}_0^2-\frac{1}{\rho^2}
\int_0^{\infty}\upd\omega'\, \frac{{\omega'}^2\, v_{\omega'}^2}
{(\omega+i0)^2-{\omega'}^2}\right]^{-1}\, . \label{4.4}
\end{equation}
Having solved the algebraic equations of the set (\ref{3.9})--(\ref{3.14})
we now turn to the differential equation (\ref{3.10}). 
\section{General form of the tensorial coefficients}
After elimination of $\tens{f}_{A}$ and $\tens{f}_{P}$ with the help of
(\ref{3.9}) and (\ref{3.11}) we get from (\ref{3.10}):
\begin{equation}
\Delta\, \tens{f}_{\Pi}({\bm r},{\bm r}',\omega)+\frac{\omega^2}{c^2}
\tens{f}_{\Pi}({\bm r},{\bm r}',\omega)=-\frac{\omega^2}{c^2}\, 
\left[\alpha \tens{f}_{X}({\bm r},{\bm r}',\omega)\right]_\mathrm{T}\,
. \label{5.1}
\end{equation}
Since $\tens{f}_{\Pi}({\bm r},{\bm r}',\omega)$ is purely transverse in
${\bm r}$, we may write the first term at the left-hand side as $-{\bm
\nabla}\times [{\bm \nabla}\times \tens{f}_{\Pi}]$. Subsequently, we
introduce $\tens{f}_{E}$ by means of (\ref{3.5}), and eliminate
$\tens{f}_{X}$ with the help of (\ref{4.3}). The result is an
inhomogeneous wave equation for $\tens{f}_{E}$:
\begin{equation}
-{\bm \nabla}\times [{\bm \nabla}\times \tens{f}_{E}({\bm r},{\bm
 r}',\omega)]+\frac{\omega^2}{c^2}\, \left[1+\chi({\bm r},\omega)\right]\, 
\tens{f}_{E}({\bm r},{\bm r}',\omega)=
-\frac{\omega^2}{\rho\alpha c^2}\, v_{\omega}\, \chi({\bm r},\omega)\, 
\tens{s}({\bm r},{\bm r}',\omega)\, . \label{5.2}
\end{equation}
To solve it we introduce the tensorial Green function $\tens{G}$ of this
wave equation, which is defined as:
\begin{equation}
-{\bm \nabla}\times [{\bm \nabla}\times \tens{G}({\bm r},{\bm
 r}',\omega)]+\frac{\omega^2}{c^2}\, \left[1+\chi({\bm r},\omega)\right]\, 
\tens{G}({\bm r},{\bm r}',\omega)=\tens{I}\, \delta({\bm r}-{\bm r}')\,
 . \label{5.3}
\end{equation}
In terms of $\tens{G}$ the solution of (\ref{5.2}) reads:
\begin{equation}
\tens{f}_{E}({\bm r},{\bm r}',\omega)=-\frac{\omega^2}{c^2}
\int\upd {\bm r}''\, \frac{1}{\rho''\alpha'' }\, v''_{\omega}\, 
\chi({\bm r}'',\omega)\, \tens{G}({\bm r},{\bm r}'',\omega)\cdot
\tens{s}({\bm r}'',{\bm r}',\omega)\, . \label{5.4}
\end{equation}
Now that we have found $\tens{f}_{E}$ in terms of $\tens{s}$ it is
straightforward to express all other tensorial coefficients in
$\tens{s}$. We first list the results for the coefficients of the field and
polarization variables:
\begin{eqnarray}
\tens{f}_{A}({\bm r},{\bm r}',\omega)&=&
-\frac{i}{\omega}\, 
\left[\tens{f}_{E}({\bm r},{\bm r}',\omega)\right]_\mathrm{T}\, , 
\label{5.5}\\
\tens{f}_{\Pi}({\bm r},{\bm r}',\omega)&=&
-\varepsilon_0\, \left[\tens{f}_{E}({\bm r},{\bm r}',\omega)\right]_\mathrm{T}
\, , \label{5.6}\\
\tens{f}_{X}({\bm r},{\bm  r}',\omega)&=& 
-\frac{\varepsilon_0}{\rho\alpha^2}\, v_{\omega} \chi({\bm r},\omega)\, 
\tens{s}({\bm r},{\bm  r}',\omega)
-\frac{\varepsilon_0}{\alpha}\, \chi({\bm r},\omega)\, 
\tens{f}_{E}({\bm r},{\bm r}',\omega)\, , \label{5.7}\\
\tens{f}_{P}({\bm r},{\bm  r}',\omega)&=& 
\frac{i\varepsilon_0}{\alpha^2}\, \omega\, v_{\omega} \chi({\bm r},\omega)\, 
\tens{s}({\bm r},{\bm  r}',\omega)
+i\alpha\, \frac{1}{\omega}\,  
\left[\tens{f}_{E}({\bm r},{\bm r}',\omega)\right]_\mathrm{T}\nonumber\\
&&+\frac{i\varepsilon_0\rho}{\alpha}\, \omega\, \chi({\bm r},\omega)\,
\tens{f}_{E}({\bm r},{\bm r}',\omega)\, , \label{5.8}
\end{eqnarray}
where (\ref{5.4}) should be inserted. The coefficients for the bath
variables are
\begin{eqnarray}
\tens{f}_{Y}({\bm r},{\bm  r}',\omega',\omega)&=&
\frac{i}{\rho\omega}\, \tens{s}({\bm r},{\bm r}',\omega)\, 
\delta(\omega-\omega')
-\frac{i\varepsilon_0}{\rho^2\alpha^2}\, 
\frac{\omega v_{\omega}  v_{\omega'}}{(\omega+i0)^2-{\omega'}^2}\, 
\chi({\bm r},\omega)\, \tens{s}({\bm r},{\bm r}',\omega)\nonumber\\
&&-\frac{i\varepsilon_0}{\rho\alpha}\, 
\frac{\omega v_{\omega'}}{(\omega+i0)^2-{\omega'}^2}\, \chi({\bm r},\omega)\, 
\tens{f}_{E}({\bm r},{\bm r}',\omega)\, , \label{5.9}\\
\tens{f}_{Q}({\bm r},{\bm  r}',\omega',\omega)&=&
-i\rho\frac{{\omega'}^2}{\omega}\, \tens{f}_{Y}({\bm r},{\bm  r}',\omega',\omega)\,
. \label{5.10}
\end{eqnarray}
The expressions listed here still depend on the tensor $\tens{s}$, which is
not yet known. It can be determined by employing the commutation relations
(\ref{3.2}).
\section{Determination of the tensor $\tens{s}({\bm r},{\bm r}',\omega)$}
The general form (\ref{3.7}) should satisfy the commutation relations
(\ref{3.2}). The first of these leads upon using eqs.~(\ref{2.2}) to the
following constraint:
\begin{eqnarray}
&&\rule{-0.5cm}{0cm}\int \upd{\bm r}''  \biggl\{  
\tens{f}_{A}^{\dagger}({\bm r},{\bm r}'',\omega)\cdot
\tens{f}_{\Pi}({\bm r}'',{\bm r}',\omega')
-\tens{f}_{\Pi}^{\dagger}({\bm r},{\bm r}'',\omega)\cdot
\tens{f}_{A}({\bm r}'',{\bm r}',\omega')\nonumber\\
&&+\tens{f}_{X}^{\dagger}({\bm r},{\bm r}'',\omega)\cdot
\tens{f}_{P}({\bm r}'',{\bm r}',\omega')
-\tens{f}_{P}^{\dagger}({\bm r},{\bm r}'',\omega)\cdot
\tens{f}_{X}({\bm r}'',{\bm r}',\omega')\nonumber\\
&&+\int_0^{\infty}\upd\omega''\biggl[ 
\tens{f}_{Y}^{\dagger}({\bm r},{\bm r}'',\omega'',\omega)\cdot
\tens{f}_{Q}({\bm r}'',{\bm r}',\omega'',\omega')
-\tens{f}_{Q}^{\dagger}({\bm r},{\bm r}'',\omega'',\omega)\cdot
\tens{f}_{Y}({\bm r}'',{\bm r}',\omega'',\omega')\biggr]\biggr\}\nonumber\\
&&=-i\hbarit\, \delta(\omega-\omega')\, \tens{I}\, \delta({\bm r}-{\bm r}')\,
, \label{6.1}
\end{eqnarray}
where $\tens{f}^{\dagger}({\bm r},{\bm r}')$ equals
$\tens{f}^{\ast}({\bm r}',{\bm r})$ with interchanged tensor indices. If
the expressions (\ref{5.5})--(\ref{5.10}) are substituted at the left-hand
side, we find contributions with either no factor $\tens{s}$, or one or two
such factors. We first consider the contribution that depends quadratically
on $\tens{s}$. In the ${\bm r}''$-integrand in (\ref{6.1}) one encounters a
frequency integral, which can be rewritten in terms of $\chi({\bm
r}'',\omega)$ and $\chi({\bm r}'',\omega')$. After doing so, one finds that
most terms depending quadratically on $\tens{s}$ drop out. A single term of
this type survives. It yields the following contribution to the left-hand
side of (\ref{6.1}):
\begin{equation}
-\frac{2i}{\omega}\, \delta(\omega-\omega')\, 
\int\upd{\bm r}''\, \frac{1}{\rho''}\, \tens{s}^\dagger({\bm r},{\bm
  r}'',\omega)\cdot\tens{s}({\bm r}'',{\bm r}',\omega)\, . \label{6.2}
\end{equation}
The remaining contributions at the left-hand side of (\ref{6.1}) can
likewise be evaluated by first rewriting the frequency integrals in terms
of the susceptibility. Subsequently, one finds that the integrands contain
terms with the transverse part $\left[\tens{f}_{E}^{\dagger}({\bm r},{\bm
r}'',\omega)\right]_{\mathrm{T}''}$ of $\tens{f}_{E}^{\dagger}({\bm r},{\bm
r}'',\omega)$. Upon adding all such terms one establishes that this
transverse part is multiplied by the combination
\begin{equation}
[1+\chi({\bm r}'',\omega')]\, 
\tens{f}_{E}({\bm r}'',{\bm r}',\omega')
+\frac{1}{\rho''\alpha''}\, v''_{\omega'}\, \chi({\bm r}'',\omega')\, 
\tens{s}({\bm r}'',{\bm r}',\omega')\, . \label{6.3}
\end{equation}
This combination is itself purely transverse in ${\bm r}''$, as one proves
directly from (\ref{5.2}). Hence, in the integral over ${\bm r}''$ one may
replace the transverse part $\left[\tens{f}_{E}^{\dagger}({\bm r},{\bm
r}'',\omega)\right]_{\mathrm{T}''}$ by the full coefficient 
$\tens{f}_{E}^{\dagger}({\bm r},{\bm r}'',\omega)$. A similar remark
applies to the terms with the transverse part of $\tens{f}_{E}({\bm
r}'',{\bm r}',\omega')$. After these replacements, one arrives at an
integral expression, which after use of (\ref{5.2}) becomes proportional to
\begin{equation}
\int\upd{\bm r}''
\biggl[
\biggl\{{\bm \nabla}''\times [{\bm \nabla}''\times 
\tens{f}_{E}^{\dagger}({\bm r},{\bm r}'',\omega)]\biggr\}\cdot 
\tens{f}_{E}({\bm r}'',{\bm r}',\omega')
-\tens{f}_{E}^{\dagger}({\bm r},{\bm r}'',\omega)\cdot
\biggl\{{\bm \nabla}''\times [{\bm \nabla}''\times
\tens{f}_{E}({\bm r}'',{\bm r}',\omega')]\biggr\}\biggr]\, . \label{6.4}
\end{equation}
A partial integration shows that this integral vanishes. Hence, the only
terms that contribute to the left-hand side of (\ref{6.1}) are those
depending quadratically on $\tens{s}$, as given in (\ref{6.2}). As a
consequence, the tensor $\tens{s}$ has to fulfill the condition:
\begin{equation}
\int\upd{\bm r}''\, \frac{1}{\rho''}\, \tens{s}^\dagger({\bm r},{\bm
  r}'',\omega)\cdot\tens{s}({\bm r}'',{\bm r}',\omega)=
\frac{\hbarit\omega}{2}\, \tens{I}\, \delta({\bm r}-{\bm r}')\, . \label{6.5}
\end{equation}
Hence, the general form for $\tens{s}$ is
\begin{equation}
\tens{s}({\bm r},{\bm r}',\omega)=\sqrt{\frac{\hbarit\omega\rho}{2}}\, 
\tens{U}({\bm r},{\bm r}',\omega)\, . \label{6.6}
\end{equation}
The frequency-dependent tensor $\tens{U}({\bm r},{\bm r}',\omega)$ must be
unitary, so that it has the property $\int\upd{\bm r}''\,
\tens{U}^{\dagger}({\bm r},{\bm r}'',\omega)\cdot \tens{U}({\bm r}'',{\bm
r}',\omega) =\tens{I}\, \delta({\bm r}-{\bm r}')$. One may check that the
second commutation relation in (\ref{3.2}) does not lead to new constraints
on $\tens{s}$.

Upon inserting (\ref{6.6}) and (\ref{5.4}) in (\ref{5.5})--(\ref{5.10}),
and substituting the latter in (\ref{3.7}), we obtain the diagonalizing
operators ${\bm C}({\bm r},\omega)$ and ${\bm C}^{\dagger}({\bm
r},\omega)$. They are determined up to a unitary transformation, which
leaves both (\ref{3.1}) and (\ref{3.2}) invariant. The simplest choice for
$\tens{U}$ is a local diagonal tensor $\exp[i\psi({\bm r},\omega)]\,
\tens{I}\, \delta({\bm r}-{\bm r}')$, with an arbitrary phase factor
$\exp[i\psi({\bm r},\omega)]$. A convenient form for this phase factor is
$\exp[i\psi({\bm r},\omega)]=i\, \chi^{\ast}({\bm r},\omega)/|\chi({\bm
r},\omega)|$, as it enables us to eliminate $v_{\omega}$ from the tensorial
coefficient (\ref{5.4}). In fact, from (\ref{4.4}) one can prove the
identity $v_\omega=\alpha\, [2\rho\, \mathrm{Im}\chi({\bm
r},\omega)/(\pi\varepsilon_0 \omega)]^{1/2}/|\chi({\bm r},\omega)|$.

Now that the diagonalizing operators of the model have been found, we can
switch to the Heisenberg picture and derive explicit expressions for the
time-dependent operators representing the electric field and the
displacement field. With the above choice for the unitary transformation and
the associated phase factor we obtain:
\begin{eqnarray}
{\bm E}({\bm r},t)&=&-i\mu_0 \int 
\upd{\bm r}' \int_0^{\infty} \upd \omega\, \omega
\, e^{-i\omega t}\, \tens{G}({\bm r},{\bm r}',\omega) \cdot
{\bm J}({\bm r}',\omega) +\tx{h.c.}\, , \label{6.7}\\
{\bm D}({\bm r},t)&=&-\frac{i}{c^2}\int \upd{\bm  r}'
\int_0^{\infty}\upd\omega\, e^{-i\omega t}\, \omega \, 
[1+\chi({\bm r},\omega)]\, \tens{G}({\bm r},{\bm r}',\omega)\cdot
{\bm J}({\bm r}',\omega)\nonumber\\
&&+i\int_0^{\infty} \upd\omega\, e^{-i\omega t}\, \frac{1}{\omega}\, 
{\bm  J}({\bm r},\omega)+\tx{h.c.}\, , \label{6.8}
\end{eqnarray}
with a current density ${\bm J}$ that is proportional to the
diagonalizing operator:
\begin{equation}
{\bm J}({\bm r},\omega)=\sqrt{\frac{\hbarit\varepsilon_0\,\mathrm{Im}\chi({\bm
r},\omega)}{\pi}}\, \omega\, {\bm C}({\bm r},\omega)\, . \label{6.9}
\end{equation}
Similar expressions can be found for the Heisenberg operators that
represent other canonical variables. 

The expressions (\ref{6.7}) and (\ref{6.8}) agree with those found before
by means of a Laplace-transform technique~\cite{SW04}. The electric field
${\bm E}({\bm r},t)$ at the position ${\bm r}$ and the time $t$ is given by
an integral transform, which involves the current density ${\bm J}({\bm
r}',\omega)$ as a source and the Green function $\tens{G}({\bm r},{\bm
r}',\omega)$ propagating a disturbance from ${\bm r}'$ to ${\bm r}$. The
displacement field ${\bm D}({\bm r},t)$ is the sum of two terms. First, it
contains a contribution that is closely analogous to the expression for the
electric field ${\bm E}({\bm r},t)$, with the local permeability
$1+\chi({\bm r},\omega)$ as an extra factor in the integral transform. The
second term in the polarization density depends on the local current
density ${\bm J}({\bm r},\omega)$ only, without an intervening
permeability, and without propagation effects. For the homogeneous case the
expressions (\ref{6.7}) and (\ref{6.8}) reduce to those found
in~\cite{HB92}. In a phenomenological quantization scheme~\cite{GW959698},
forms like (\ref{6.7}) and (\ref{6.8}) are postulated without proof. In
that context ${\bm J}({\bm r},\omega)$ is called a noise-current density,
which is introduced without giving its precise connection to the
fundamental dynamical variables of the system. In contrast, the present
model leads to the specific expression (\ref{6.9}) with (\ref{3.7}) for
${\bm J}({\bm r},\omega)$.

\section{Conclusion}
By using Fano's diagonalization procedure we have succeeded in finding the
ladder operators ${\bm C}$ and ${\bm C}^{\dagger}$, which annihilate and
create the normal-mode excitations of the inhomogeneous damped-polariton
model. Knowledge of these fundamental operators suffices to determine the
full time dependence of the dynamical operators that describe the field and
the dielectric. The general structure of these time-dependent operators
agrees with that postulated in a phenomenological quantization scheme.
Hence, the present diagonalization of the inhomogeneous damped-polariton
model gives a justification of the phenomenological quantization method for
inhomogeneous absorptive dielectrics.

\end{document}